\documentclass[english,showpacs,twocolumn,aps,pra,reprint,superscriptaddress]{revtex4-1}

\usepackage{mathptmx}
\usepackage[T1]{fontenc}
\usepackage[latin9]{inputenc}
\usepackage{color}
\usepackage{amsmath}
\usepackage{amssymb}
\usepackage{hyperref}
\usepackage{graphics}
\usepackage{graphicx}
\usepackage{epsfig}
\usepackage{bm}
\usepackage{epstopdf}
\usepackage{mathtools}
\usepackage{soul}
\usepackage{subfigure}   
\usepackage{xmpmulti}
\usepackage{lipsum}

\usepackage{array}
\usepackage{tabularx}
\usepackage{multirow}
\usepackage[thinlines]{easytable}

\def\beq{\begin{equation}}
\def\eeq{\end{equation}}

\def\reff#1{(\ref{#1})}
\def\rhoc{\rho_\mathrm{c}}
\def\rhoi{\rho_\mathrm{i}}

\def\Up{U_\mathrm{p}}

\def\Wcmcm{\mbox{\rm W/cm$^{2}$}}

\def\omegaM{\omega_\mathrm{M}}

\def\omegaC{\Omega_\mathrm{c}}
\def\omegaC0{\Omega_\mathrm{c0}}

\def\vekt#1{\bm{#1}}

\def\pabl#1#2{\frac{\partial #1}{\partial #2}}

\def\v0{v_0}
\def\I0{I_0}

\definecolor{dgreen}{rgb}{0.0, 0.5, 0.0}
\definecolor{dblue}{rgb}{0.0, 0.0, 0.5}
\definecolor{green}{rgb}{0.0,0.5,0.0}

\usepackage{babel}

\begin{document}

\title{Laser cluster interaction in external magnetic field: emergence of nearly mono-energetic weakly relativistic electron beam}

\author{Kalyani Swain}
\affiliation{Institute for Plasma Research, Bhat, Gandhinagar, 382428, India}
\affiliation{Homi Bhabha National Institute, Training School Complex, Anushaktinagar, Mumbai 400094, India}

\author{S. S. Mahalik}
\thanks{Presently at Bellatrix Aerospace Pvt. Ltd., Bangalore, 560020, India}
\affiliation{Institute for Plasma Research, Bhat, Gandhinagar, 382428, India}

\author{M. Kundu}
\affiliation{Institute for Plasma Research, Bhat, Gandhinagar, 382428, India}
\affiliation{Homi Bhabha National Institute, Training School Complex, Anushaktinagar, Mumbai 400094, India}

\date{\today}
\begin{abstract}
Recent studies [Sci Rep 12, 11256 (2022)] on laser interaction (wavelength 800~nm, intensity $>10^{16}\, \Wcmcm$) with deuterium nano-cluster in an ambient magnetic field ($B_0$) demonstrate that collisionless absorption of laser occurs in two stages via anharmonic resonance (AHR) and electron-cyclotron resonance (ECR) or relativistic ECR (RECR) processes. Auxiliary $B_0$ enhances coupling of laser to cluster-electrons via improved frequency-matching for ECR/RECR as well as phase-matching for prolonged duration of the 5-fs (fwhm) broadband pulse and the average absorbed energy per electron $\overline{\mathcal{E}}_A$ significantly jumps up $\approx 36-70$ times of its ponderomotive energy ($\Up$). In this paper, we report energy dispersion of these energetic electrons and their angular distribution in position and momentum space by performing hybrid-PIC simulations. 
By simulating bigger clusters (radius $R_0 \approx 3-4$~nm) at high intensities $\approx 10^{16} - 10^{18}\,\Wcmcm$, we find $\overline{\mathcal{E}}_A\approx 36-70\,\Up$ similar to a small cluster ($R_0\approx 2$~nm), but total energy absorption increases almost linearly with increasing cluster size due to more number of available energy carriers. 
And, in all cases (near ECR/RECR) electrons are collimated like a weakly relativistic gyrating beam (about $B_0$) within an angular spread $\Delta\theta<5^{\circ}$, propagating far beyond $200 R_0$ along $B_0$. This study may be relevant because an intense, collimated electron beam has wide applications including the fast ignition technique for inertial confinement fusion, ultra-short x-ray sources and medical applications. 

\end{abstract}

\maketitle
\section{Introduction}\label{sec1}
The interaction of intense laser field with cluster of atoms or molecules (called nano-clusters) constitutes a promising research area in strong-field wave-matter interaction. Atomic-clusters, with locally high atom-density resembling those in a solid, may absorb $80-90\%$ of laser compared to laser-solid and laser-atomic gas interactions\cite{Ditmire_PRL78_2732}. Importantly, laser-cluster interaction (LCI) produces energetic ions\cite{Ditmire_PRL78_2732,Ditmire_PRA57,Ditmire_Nature386,Lezius}, neutrals\cite{Rajeev_Nature}, electrons\cite{Ditmire_PRA57,Chen_POP_9,Shao_PRL77,Springate_PRA68,Chen_PRE_2002} and x-rays\cite{McPherson_Nature370,Chen_PRL104,Jha_2005,Dorchies_Xray,Kumarappan2001}, and thus paves the way to future generation particle accelerators and photon accelerators. Basic processes in LCI: (i) inner ionization ---birth of electrons leading to formation of nano-plasma, (ii) outer ionization ---removal electrons from the whole cluster, (iii) coulomb explosion ---acceleration of background ions are described elsewhere\cite{RosePetruck,Bauer2003,Siedschlag_PRA_67_2003,Snyder_PRL_1996_IG}, and not repeated here for the sake of conciseness.


For laser intensities $\I0>10^{16}\,\Wcmcm$ and wavelength $\lambda>600$~nm, laser absorption in cluster is mostly collision-less\cite{Ishikawa,Megi,Jungreuthmayer_2005,Bauer2004} wherein linear resonance (LR) and anharmonic resonance (AHR) may play active roles. LR happens\cite{Ditmire_PRA53,LastJortner1999,Saalmann2003,Fennel_EPJD_29_2004} for a long duration laser pulse (typically > 50 fs) on coulomb explosion of an initially over-dense ($\rhoi>\rhoc$) cluster when ionic charge density $\rhoi(t)$ gradually drops to the critical density $\rhoc = \omega^2/4\pi$ and the Mie-plasma frequency $\omegaM(t)=\sqrt{4\pi \rhoi/3}$ dynamically meets the laser frequency $\omega=2\pi c/\lambda$. Atomic units (a.u.) $\vert e\vert = m_0 =  \hslash = 4\pi\epsilon_0 = 1$ are used unless noted explicitly.  However, for short pulse duration $\omegaM(t)>\omega$ holds (cluster is over-dense) and AHR becomes important. During AHR, oscillation frequency of an electron in the self-consistent ahharmonic cluster-potential meets $\omega$. AHR is noted in many works\cite{MulserPRA,MulserPRL,MKunduPRA2006,MKunduPRL,Kostyukov,Taguchi_PRl,SagarPOP2016} using rigid sphere model (RSM), particle-in-cell (PIC) and molecular dynamics (MD) simulations. 

Though numerous experiments, analytical models and numerical simulations have shown energetic electrons in the collisionless regime, our survey\cite{Swain2022} reveals that maximum average energy of a liberated electron mostly remains close to $\approx 3.17\Up$ similar to the laser-atom\cite{MorenoEPL_1994,MorenoPRA_1997,MLeinPRL2003} interaction ($\Up = I_0/4\omega^2$ is the ponderomotive energy of an electron) or below; {\em even if} laser has adequate supply of energy. In case of laser-deuterium cluster interaction\cite{Swain2022}, with a 5-fs broadband laser pulse (central $\lambda = 800$~nm, $\I0 \approx 10^{16} - 10^{18}\,\Wcmcm$) and an ambient magnetic field ($B_0$) in crossed orientation to the laser electric field ($E_l$), earlier we have shown by RSM and three dimensional PIC simulation that enhanced laser absorption occurs in two stages via AHR ($1^{st}$ stage) and electron cyclotron resonance (ECR) or relativistic ECR (RECR) processes ($2^{nd}$ stage). During ECR/RECR, electron cyclotron frequency $\Omega_\mathrm{c}=\vert e \vert B_0/m_0\gamma = \omegaC0/\gamma$ meets~$\omega$. The auxiliary $B_0$ enhances coupling of laser to cluster-electrons via improved frequency-matching as well as phase-matching and the average absorbed energy per electron jumps to $\overline{\mathcal{E}}_A\approx 36-70\,\Up$ (more than $12-36$ fold\cite{Swain2022}) which {\em is significant}.

One may argue that required ambient $B_0 =10-20$ kT is too high to achieve above ECR/RECR in a laboratory with $800$~nm laser. With~$\mathrm{CO_2}$ laser ($\lambda \approx 10.6\, \mu\mathrm{m}$), however, the strength of $B_0$ for the ECR/RECR is scaled down to $B_{0} \approx 1-2$ kT which seems to be feasible (and our simulations are underway).
Recent demonstration of pulsed magnetic fields from sub kilo-Tesla \cite{Shaikh_2016,Ivanov_2021} and kilo-Tesla to mega-Tesla \cite{Fujioka2013,Nakamura_2018, Murakami2020,Wilson_2021,Longman_2021} has renewed interest in laser-plasma \cite{Shi_appl_laserplasma,Gong_2020,Weichman_2020} community and may serve the purpose.
In this context, we mention that self-generated (quasi-static) magnetic fields beyond 10~kT are noted in high density laser-plasma experiments and astro-physical conditions. For example, self-generated magnetic fields $\approx 20-46$~kT have been measured \cite{Tatarakis2002,Tatarakis_pop2002}. 
Magnetic fields around neutron stars and pulsars\cite{ShapiroAndTeukolsky} typically vary $\approx 10^1-10^5$~kT. Understanding of the origin of energetic electrons in these strong electromagnetic field conditions are also of fundamental interest. 
%
From the application point of view, energetic electrons produced by LCI via ECR/RECR in an ambient magnetic field can be helpful for the table-top radiation sources (such as x-rays), particle-accelerators useful for medical applications and inertial confinement fusion (ICF).

In the previous work\cite{Swain2022}, though it is shown that average energy of laser-driven cluster-electrons increases significantly ($\overline{\mathcal{E}}_A\approx 36-70\,\Up$) with an ambient $B_0$ near ECR/RECR, it is not yet known how those electrons propagate. Generation of relativistic electron beam (REB) is also of current interest\cite{Iwawaki,Malko2019,malkov_2013}. 
Therefore, understanding the energy distribution of ejected cluster-electrons and their divergence (directional) properties are important from the point of view of applications as well as in the astrophysical scenario mentioned above. In this paper, we report energy dispersion of these energetic electrons and their angular distribution in position and momentum space by performing hybrid-PIC simulations. The effect of ambient magnetic field-driven ECR/RECR on different cluster size is also not known so far. This may be particularly important for higher electron flux as a collimated beam. By simulating relatively bigger clusters of radius $R_0 \approx 3-4$~nm at intensities $\I0 \approx 10^{16} - 10^{18}\,\Wcmcm$, here we find that average absorbed energy per electron jumps to $\overline{\mathcal{E}}_A\approx 36-70\,\Up$ similar to a small cluster ($R_0\approx 2$~nm), but total energy absorption increases almost linearly with increasing cluster size due to more number of available energy carriers. 
And, in all cases (near ECR/RECR) electrons are collimated like a weakly relativistic gyrating beam (about $B_0$) within an angular spread $\Delta\theta<5^{\circ}$, propagating far beyond $200 R_0$ along $B_0$. 

In Sec.\ref{secPulse} laser pulse and cluster parameters are given. In Sec.\ref{sec2} we discuss details of the PIC simulation code and its new hybrid capability for treating particle-interactions outside the simulation box. Section \ref{sec3} focuses on the energy distribution of electrons and their angular distribution in the position space (as well as in the momentum space) corresponding to the laser energy absorption by electrons for different ambient $B_0$. In Sec.\ref{sec4}, laser energy absorption for bigger clusters and associated angular distributions of ejected electrons are compared at high intensities. Sec.\ref{sec6} concludes this work.

\section{Laser pulse and cluster parameters}\label{secPulse}
We assume a laser pulse \cite{MKunduPRA2012,SagarPRA2018,Swain2022} of vector potential
$\vekt{A}_l(t') = \vekt{\hat{x}} ({E_0}/{\omega})\sin ^2({\omega t'}/{2n}) \cos (\omega t')
$ for $0 \leq t' \leq nT$
which is polarized in $\vekt{x}$ and propagating in $\vekt{z}$; where $t' = t - {z}/c$, 
%
%
$n=$ number of laser period $T$, 
$\tau = n T$ is pulse duration, and $E_0=\sqrt{I_0}$ is field strength.
Laser electric and magnetic fields $\vekt{E}_l$, $\vekt{B}_l$ read
\begin{equation}\label{eq:laserfieldE}
%
\vekt{E}_l (t')  = -\pabl{\vekt{A}_l}{t}, \,\,\,\,\,\,\,\,
\vekt{B}_l (t')  = \hat{\mathbf{z}} \times \vekt{E}_l (t') / c
\end{equation}

\noindent
The broad-band nature of the pulse is understood from its discrete frequencies $\omega_1 = \omega, \omega_2 = (1+1/n)\omega$, $\omega_3 = (1-1/n)\omega$ and side-bands are significant for short pulses.  

Deuterium clusters of different sizes and number of atoms $N = 2176, 7208, 17256$ are chosen. According to the Wigner-Seitz radius $r_w\approx 0.17$~nm, respective cluster radii are $R_0=r_w N^{1/3} \approx 2.2, 3.3, 4.4$~nm. For $R_0\ll\lambda$, the dipole approximation $z/\lambda \ll 1$ may be assumed.
Single cluster is illuminated by the above laser pulse of $\lambda = 800$~nm for $n=5$, $\tau = nT\approx 13.5$~fs ($\tau_{fwhm}\sim 5$~fs). A cluster is $\rho_i/\rhoc \approx 27.1$ times overdense with $(\omegaM/\omega)^2 \approx 9.1$, where $\rhoc \approx 1.75\times 10^{27} m^{-3}$ is the critical density at $\lambda = 800$~nm.

\section{Particle-in-cell (PIC) simulation}\label{sec2}

We use 3D PIC simulation code \cite{MKunduPRL,MKunduPRA2006,MKunduPRA2007,Popruzhenko2008,MKunduPOP2008,MKunduPRA2012,MKundu_Thesis,Swain2022} for LCI with/without ambient magnetic field $\vekt{B}_{ext}$.
Different deuterium clusters with $N = 2176, 7208, 17256$ are placed in a cubical computational box. Center of a cluster coincides the center of the computational box.
Initially laser field $\vekt{E}_l(t)$ ionizes all neutral atoms D to D$^{+}$ (assuming over-the-barrier ionization, OBI \cite{HBethe} which is valid for $\I0>10^{15}\,\Wcmcm$) after
reaching a critical strength $E_c = \vert \vekt{E}_l(t)\vert = I_p^2(Z)/4 Z$, where $I_p(Z)$
is the ionization potential for charge state $Z=1$. 
The position and velocity of a newly born electron (after the OBI) are assumed same as the parent atom/ion conserving the momentum and energy. 
Subsequent movement of electrons and ions by the driving fields create/modify space-charge field $\vekt{E}_{sc}(\vekt{r},t) = -\vekt{\nabla}\phi(\vekt{r},t)$ and corresponding potential $\phi(\vekt{r},t)$ which are time-dependent and starts from zero.

%
A PIC electron/ion has the same charge to mass ratio of a real electron/ion. 
The equation of motion of the $j\vert k$-th PIC electron/ion ($j$ for electron and $k$ for ion) reads
%
%
%
\begin{align}\label{eom1a}
\frac{d\vekt{p}_{j\vert k}}{dt}\! &=  q_{j\vert k}\! \left[ \left(\vekt{E}_l(t) + \vekt{E}_{sc}(\vekt{r}_{j\vert k},t) \right) + \vekt{v}_{j\vert k}\times \left(\vekt{B}_l + \vekt{B}_{ext} \right)\right]
\\ \label{eom1b}
\frac{d\vekt{r}_{j\vert k}}{dt} & = \vekt{v}_{j\vert k} = \frac{\vekt{p}_{j\vert k}}{\gamma_{j\vert k} m_{j\vert k}}
\end{align}
%

%
\noindent
where $\vekt{p}_{j \vert k} = m_{j\vert k} \vekt{v}_{j\vert k} \gamma_{j\vert k}, \vekt{v}_{j\vert k}, \vekt{r}_{j \vert k}, m_{j\vert k}, q_{j\vert k}, \gamma_{j\vert k}$ are momentum, velocity, position,  mass, and charge of a PIC electron/ion and $\gamma_{j\vert k} = \sqrt{1+p_{j\vert k}^2/m_{j\vert k}^2 c^2}$ respectively. In the present case, $m_j = m_0 = 1$, $m_k = M_0 = 2\times 1386$, $q_j = -1$ and $q_k = 1$ in a.u.. 
%
Poisson's equation $\nabla^2\phi_G = -\rho_G$ is solved  
for $\phi_G$ on the numerical grid (subscript $G$ indicates grid values of potential and charge density) with time-dependent monopole boundary condition. Interpolating $\phi_G$ to the particle
position corresponding potential $\phi(\vekt{r}_{j\vert k},t)$ is obtained. Field $\vekt{E}_{sc}(\vekt{r}_{j\vert k}) = -\vekt{\nabla} \phi(\vekt{r}_{j\vert k})$ is obtained by analytical differentiation \cite{MKundu_Thesis} of interpolated $\phi(\vekt{r}_{j\vert k})$ locally at $\vekt{r}_{j\vert k}$. Equations~\reff{eom1a}-\reff{eom1b} are solved by the velocity verlet method (VVM) using laser fields \reff{eq:laserfieldE}. VVM leads to better energy conservation and less numerical heating even for a bigger $\Delta t$, particularly for the relativistically intense driving fields. Electron-ion collisions are neglected in the current work due to high field strengths.
 Total absorbed energy
$\mathcal{E}(t) = \sum_l q_l\phi_l + p_l^2/2 m_l$ is obtained by summing over kinetic energy $KE = \sum_l p_l^2/2 m_l$ and potential energy $PE = \sum_l q_l\phi_l$ of all electrons and ions. For the $n=5$-cycle pulse (used here) contribution of ion kinetic energy is small and total energy is mainly due to electrons. Final absorbed energy $ {\mathcal{E}_A} = \mathcal{E}(\tau)$ in the end the laser pulse at $\tau=nT$ is also noted.
The numerical parameters in the PIC simulation (spatial and
temporal resolution, grid size, number of PIC particles/cell etc.)
are carefully chosen for negligible artificial numerical heating. Typically, we choose $64^3$ ($128^3$ for bigger cluster) grid points (cells) with uniform cell size $\Delta x=\Delta y = \Delta z = 16$~a.u., time step  $\Delta t = 0.1$ a.u., and approximately 15 particles/cell.  

\subsection{Hybrid-PIC}
In the PIC simulation, treatment of particles crossing/leaving the boundaries of the simulation box needs a special care. Often reflecting or periodic boundary conditions are used, which preserve particles inside the simulation box and Poisson equation with appropriate boundary conditions takes care of the space-charge field $\vekt{E}_{sc}(\vekt{r}_{j\vert k})$ on an inside particle. For a finite size target (e.g., cluster) in early works we have used open boundary conditions for particles, meaning that particles which leave the simulation box are free from space-charge field $\vekt{E}_{sc}(\vekt{r}_{j\vert k})$. This assumption is tested valid for the short laser pulse by keeping simulation box size $L\approx 16-20 R_0$ (typically) beyond which $\vekt{E}_{sc}\approx 0$. It also allows particles (particularly electrons) to come back inside the box or propagate similar to direct laser acceleration (DLA) outside the box obeying \reff{eom1a}-\reff{eom1b}.     For a dense electron cloud outside the simulation box (in a strong ambient $B_0$) electron-electron interaction (repulsion) may be important for the divergence/collimation of the electron beam. Therefore, we adapt a new hybrid procedure to determine $\vekt{E}_{sc}$ on an electron: as long as it is inside the box,  $\vekt{E}_{sc}$ is solely determined by the standard PIC approach; but when it is outside the box $\vekt{E}_{sc}$ is determined by the field due to total charge (including ions and electrons) inside the simulation box (monopole field) plus the fields due all other electrons outside the simulation box as in MD simulations\cite{SagarPOP2016}. 

\section{Laser absorption in cluster in ambient magnetic fields}
\label{sec3}
\begin{figure}[h]
\includegraphics[width=0.45\textwidth]{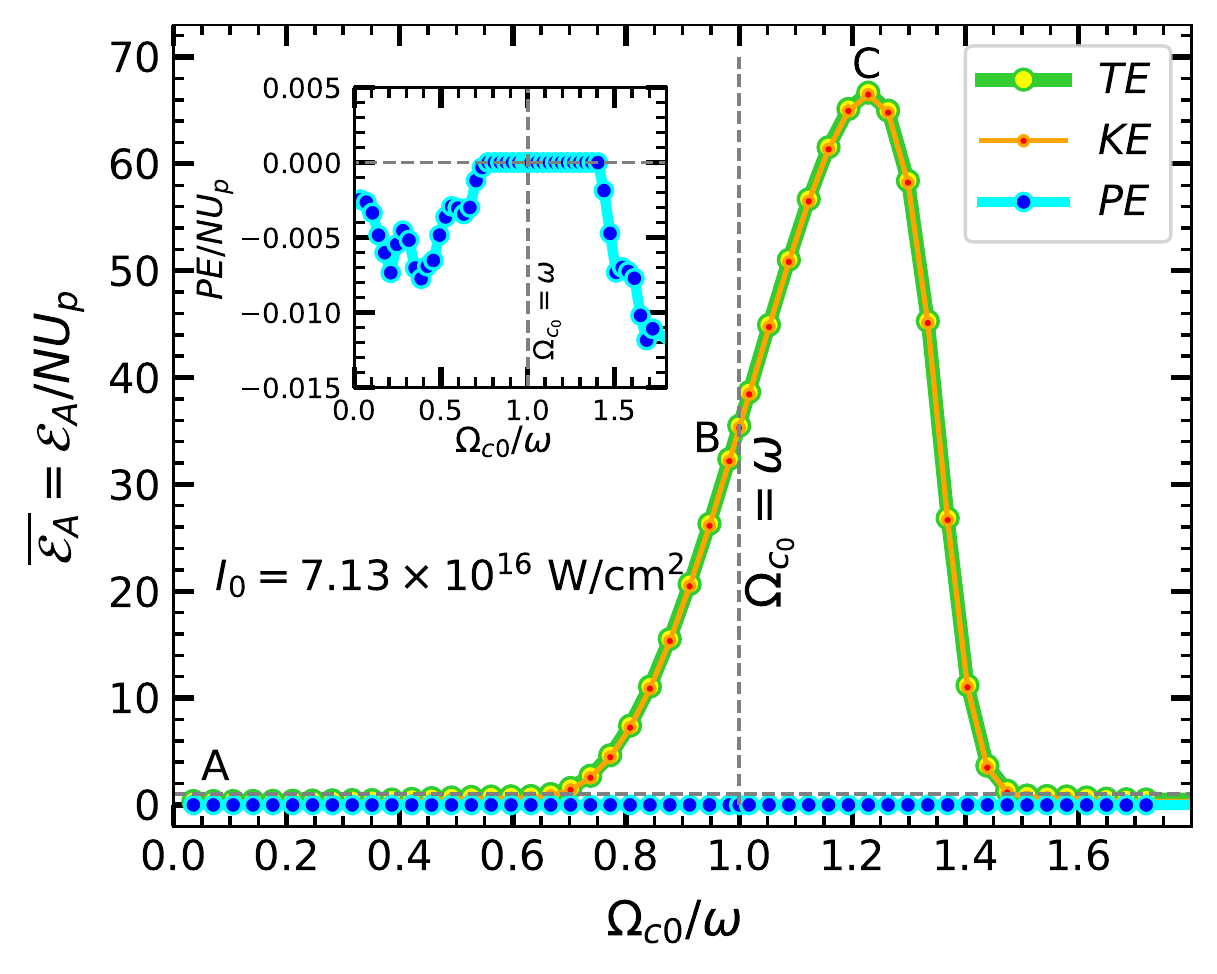}
\vspace{-0.25cm}
\caption{
PIC results: Average total energy ($TE, \overline{\mathcal{E}}_A=\overline{\mathcal{E}}(\tau)$)), kinetic energy ($KE$) and potential energy ($PE$) per cluster-electron vs normalized electron-cyclotron frequency $\omegaC0/\omega$ for a range of ambient field $\vert \vekt{\hat{z}} B_{ext}\vert\approx (0 - 2\omega)$  with $n=5$-cycle pulse of $\I0\approx 7.13\times 10^{16}\Wcmcm$ irradiating a deuterium cluster of $N = 2176$ and $R_0 = 2.2$~nm. Energy is shown normalized by corresponding $\Up$. At a low $\I0$ absorption peak occurs almost at the ECR condition $\Omega_{c0} = \omega$ (vertical dashed line, see Ref.\cite{Swain2022}). Whereas, at this high $\I0\approx 7.13\times 10^{16}\Wcmcm$, absorption is {\em even higher} and absorption peak is right-shifted from ECR condition $\Omega_{c0} = \omega$ due to relativistic modification of $\Omega_{c} = \Omega_{c0}/\gamma$ for $\gamma>1$. Absorption peak $\approx 65\Up$ give average energy per electron $\overline{{\mathcal{E}}_A}\approx 0.27$ MeV. Inset shows negligibly small average $PE$ per electron, and thus $TE$ is mainly due to $KE$.
}
\label{Fig1}
\end{figure}

A deuterium cluster of $N = 2176$ and $R_0 = 2.2$~nm is irradiated by $n=5$-cycle laser pulse of $I_0 = 7.13\times10^{16} W/cm^2$ in presence of ambient magnetic fields $\vekt{B}_{ext} = B_0\vekt{\hat{z}}$ along the laser propagation $\vekt{z}$ for different values of $B_0$.
Figure~\ref{Fig1} shows total absorbed energy $\overline{\mathcal{E}_A} = {\mathcal{E}_A}/N \Up$ per electron ($TE$, green) in units of $\Up$ vs normalized electron-cyclotron frequency $\omegaC0/\omega$. Note that $\omegaC0 = B_0$ in (a.u.). At a very low $B_0$ (or without $B_0$) absorption is very poor $\overline{\mathcal{E}_A}\approx 0.5\Up$ at the point A. It gradually increases to $\overline{\mathcal{E}_A}\approx 36 \Up$ at B for ECR (vertical dashed line, where $\omegaC0/\omega =  1$) and reaches a peak $\overline{\mathcal{E}_A}\approx 68\Up$ at C for $\omegaC0/\omega\approx 1.25$. The ratio of absorbed energies at B and C to that at A are $\overline{\mathcal{E}_A} (B)/ \overline{\mathcal{E}_A} (A) \approx 72$ and $\overline{\mathcal{E}_A} (C)/ \overline{\mathcal{E}_A} (A) \approx 136$ respectively. Thus strong ambient magnetic fields may enhance laser absorption $\approx 70-136$ folds for a cluster. Compared to laser-atom interaction\cite{MorenoEPL_1994,MorenoPRA_1997,MLeinPRL2003} where maximum energy of a liberated electron $\approx 3.17\,\Up$, these $\overline{\mathcal{E}_A}$ values in a cluster range $\approx 12-23$ folds. Though magnetic field does not work, it re-orients phase-space co-ordinates ($\vekt{r}_{j\vert k}, \vekt{v}_{j\vert k}$) of a charge particle (particularly for an electron), and hence may improve rate of laser absorption obeying the relation
\begin{align}\label{eom1d}
\frac{d({\gamma}_{j\vert k} m_{j\vert k} c^2) }{dt} & = q_{j\vert k} {\vekt{v}_{j\vert k}}.\left(\vekt{E}_l(t) + \vekt{E}_{sc}(\vekt{r}_{j\vert k},t) \right)
\end{align}
through improved phase-matching\cite{Swain2022} between $\vekt{v}_{j\vert k}$ and the total field $\vekt{E} = \vekt{E}_l(t) + \vekt{E}_{sc}(\vekt{r}_{j\vert k},t)$. Equation~\reff{eom1d} is fundamental to transfer of energy to a charge particle from the interacting fields.
Self-consistent $\vekt{E}_{sc}(\vekt{r}_{j\vert k},t)$ is nonlinear in general and falls quickly as $1/r^2$ after a few cluster radius $R_0$. AHR absorption\cite{Swain2022} of laser by an electron happens within this non-linear field and may be modified\cite{Swain2022} by an ambient $\vekt{B}_{ext}$. After coming out of the cluster via AHR (called first stage\cite{Swain2022}) with some transverse momentum, electrons are mainly controlled by remaining $\vekt{E}_l, \vekt{B}_l, \vekt{B}_{ext}$ and weak $\vekt{E}_{sc}$  outside, and there ECR/RECR may happen (called second stage\cite{Swain2022}) resulting enhanced laser absorption as in Fig.\ref{Fig1}. Results of simultaneous phase-matching and frequency-matching conditions for these absorption processes are already given in Ref.\cite{Swain2022} for similar parameters and not repeated here for the sake of conciseness. 


We also  partition total absorbed energy ($TE$, green) in electron kinetic ($KE$, red) and potential energy ($PE$, blue) in Fig.\ref{Fig1}. It shows main contribution comes form electron's kinetic energy. Acceleration of these electrons by ECR/RECR (in the second stage) resembles magnetic field assisted DLA of electrons. But, in the present case, electrons originate from the over-dense cluster, self-injected into the remaining laser field in presence of ambient $\vekt{B}_{ext}$ and no external injection mechanism is required. 
It is imortant to mention that most of models for DLA of electrons consider an under-dense, pre-formed plasma channel \cite{arefiev_robinson_khudik_2015,Pukhov_1999,Tsakiris_2000,Gong_2020,Ghotra2016a,Ghotra_2018} or single electron without considering particle interactions.  

\subsection{Energy distribution of electrons}
%

\begin{figure}[]
\includegraphics[width=0.4\textwidth]{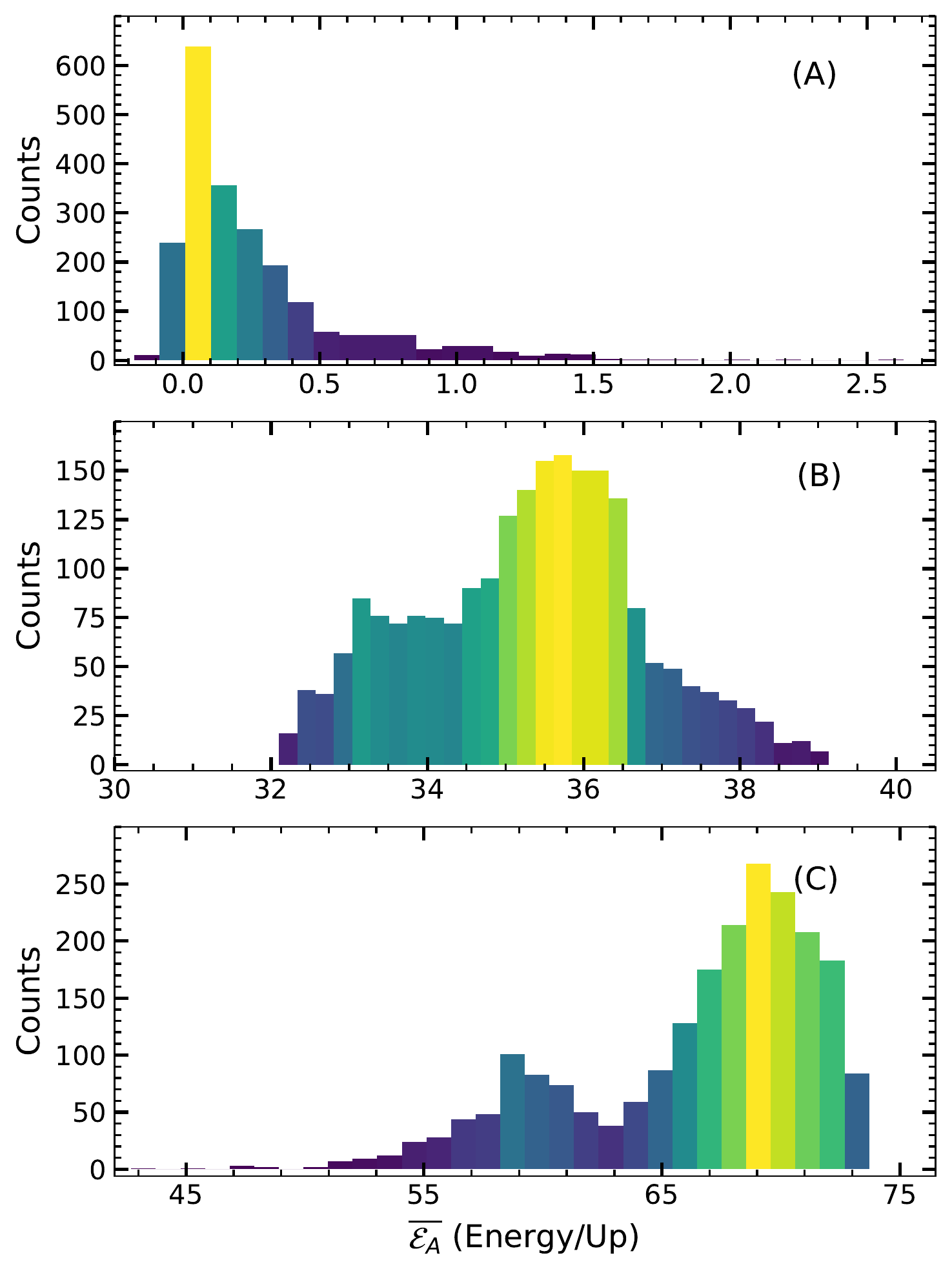}
\vspace{-0.25cm} 
\caption{Energy distribution of PIC electrons (A,B,C) ejected from the deuterium cluster corresponding to the ambient magnetic fields $B_0\approx 0.02, 0.057, 0.07$~a.u. (at A, B,C) in Fig.\ref{Fig1}. Yellow region highlights maximum electron-counts and respective energy $\overline{\mathcal{E}_A}$. Other parameters are same as in Fig.\ref{Fig1}.
}
%
%
\label{Fig2}
\end{figure}

Panels (A,B,C) in Fig.~\ref{Fig2} show energy distribution of PIC electrons corresponding to chosen data points (A,B,C) in Fig.\ref{Fig1} for $B_0 \approx 0.35\omega, \omega, 1.25\omega$ respectively. It is seen that for a given $I_0 = 7.13\times10^{16} W/cm^2$, energy distribution of electrons gradually modifies as $B_{0}$ increases. With a low value of $B_0$ (or without it, for A), more electrons (yellow region) are near lower energy $\overline{\mathcal{E}_A} \approx 0.1$, though energy tail (maximum) with a few electrons extends upto $\overline{\mathcal{E}_A} \approx 2.6$. This is the typical energy distribution of electrons one mostly finds in case of LCI with a very low $B_0$ (or without $B_0$). On the other hand, for higher $B_0$ values corresponding to B and C, more electrons are pushed around $\overline{\mathcal{E}_A} \approx 36\Up$ and $\overline{\mathcal{E}_A} \approx 68\Up$ respectively. Thus there is a reversal in the nature (variation) of energy distribution while passing from A to C.
The integrated average energy values of $\overline{\mathcal{E}_A}$ from these distributions are found to satisfy respective values of absorption at (A,B,C) in Fig.\ref{Fig1}. 
These group of electrons will now be thoroughly analyzed to understand their divergence (and collimation) as a beam.

\begin{figure}[]
\includegraphics[width=0.5\textwidth]{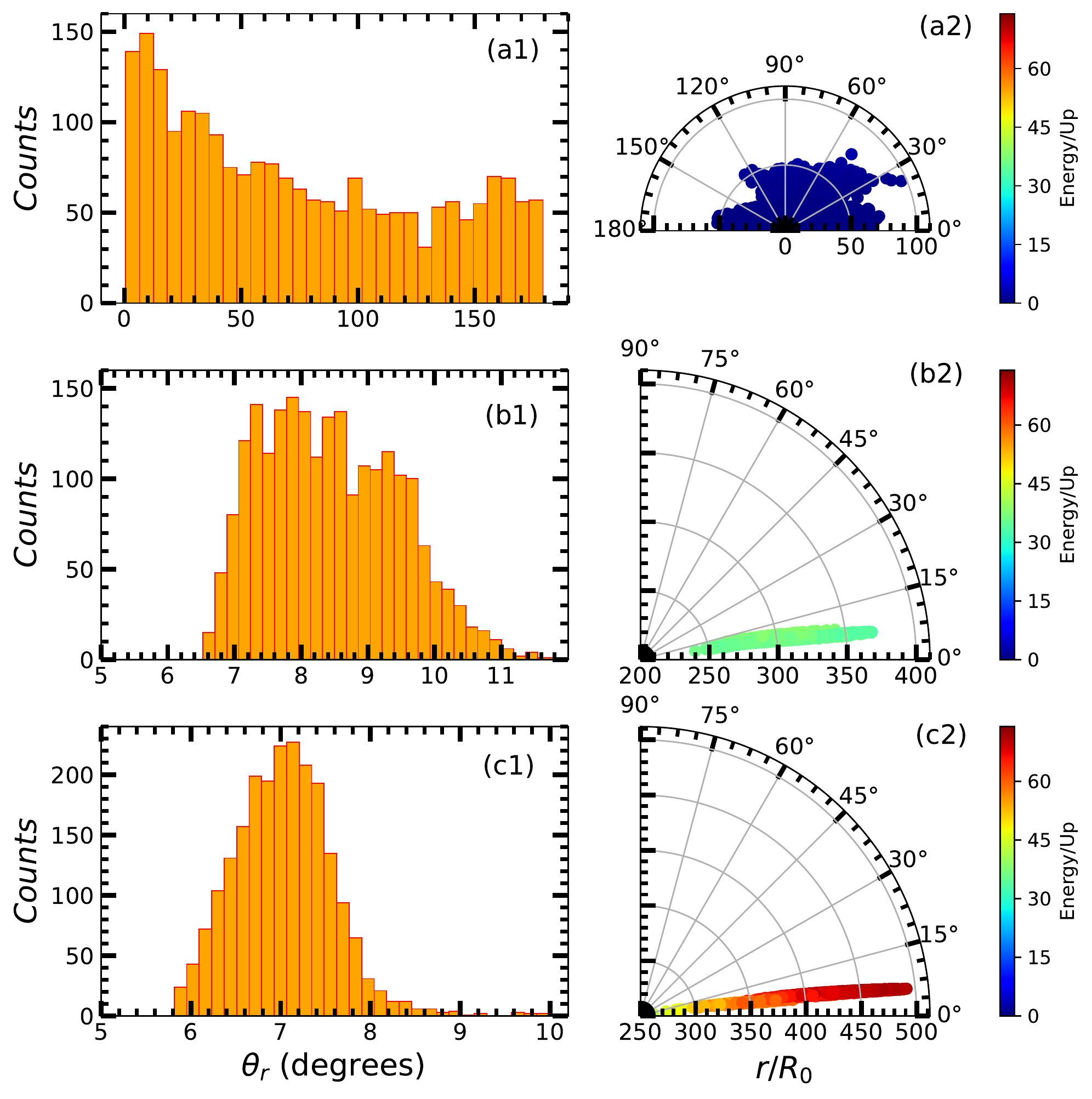}
\vspace{-0.25cm}
\caption{Histograms/distributions of angular deflection $\theta_r$ of PIC electrons (a1,b1,c1, left column) and 
respective polar plots (a2,b2,c2, right column) with their normalized position $r/R_0$ vs $\theta_r$ corresponding to those energy spectra (A,B,C) in Fig.\ref{Fig2} for $B_0=0.02, 0.057, 0.07$~a.u. respectively. Polar co-ordinates ($r,\theta_r$) are color-coded with their energy normalized by $\Up$. Ejected electrons propagate long distance $r = \sqrt{r_\perp^2 + z^2} \approx 375 R_0,500 R_0$ (b2,c2) as collimated beams with angular spreads $\Delta\theta_r<3^{\circ}$ centered around $\theta_r\approx 7-8.5^{\circ}$. Other parameters are same as in Fig.\ref{Fig1} and Fig.\ref{Fig2}.
}
\label{Fig3}
\end{figure}
\begin{figure}[]
\includegraphics[width=0.5\textwidth]{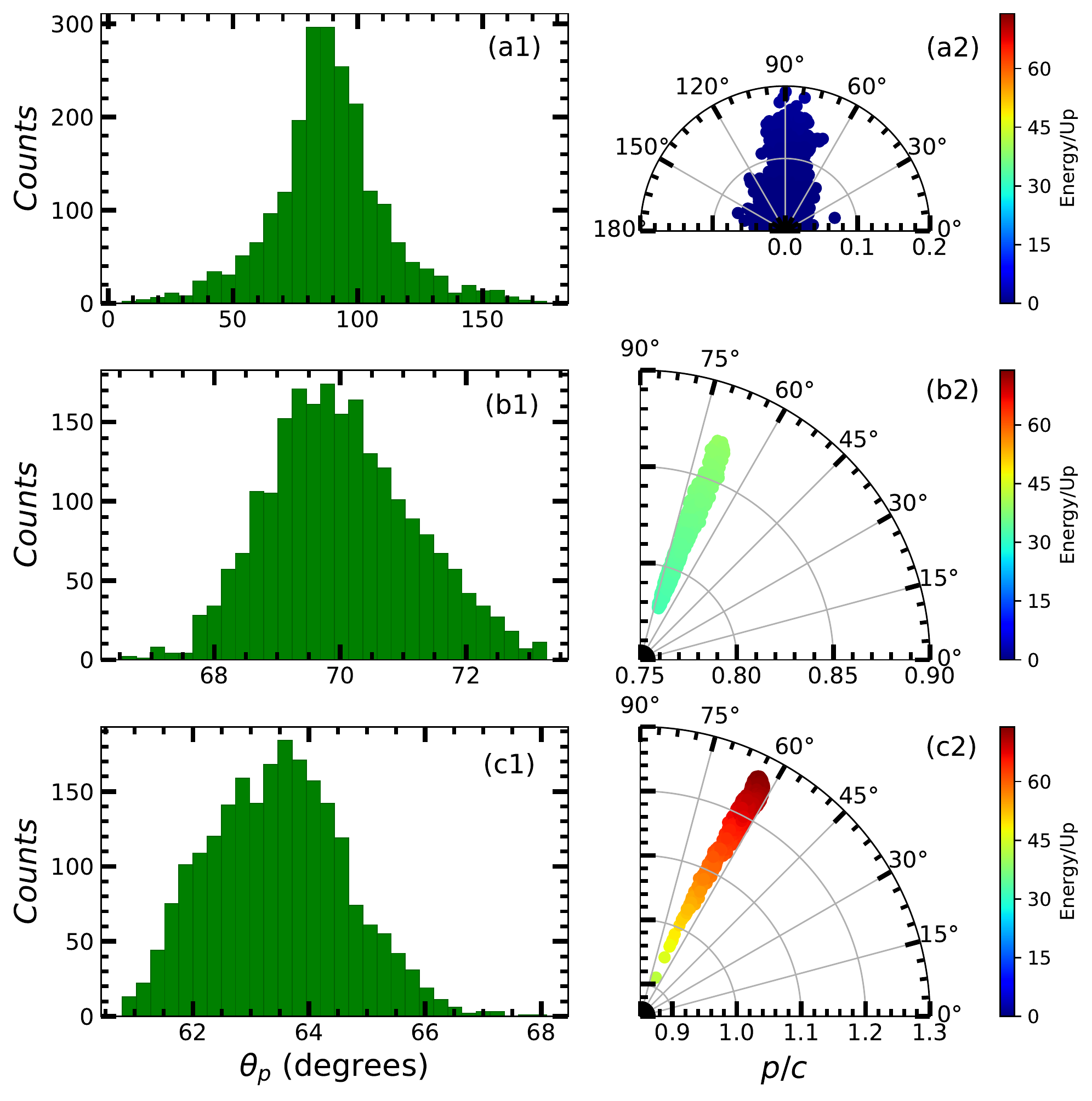}
\vspace{-0.25cm}
\caption{Histograms/distributions of angular deflection $\theta_p$ of PIC electrons (a1,b1,c1, left column) in the momentum space and 
respective polar plots (a2,b2,c2, right column) with their normalized momentum $p/c$ vs $\theta_p$ corresponding to energy spectra (A,B,C) in Fig.\ref{Fig2} for $B_0=0.02, 0.057, 0.07$~a.u. respectively. Polar co-ordinates ($p,\theta_p$) are color-coded with their energy normalized by $\Up$. Momentum of ejected electrons reach 
$p = \sqrt{p_\perp^2 + p_z^2} \approx 0.875 c,1.25 c$ (b2,c2) as collimated beams with angular spreads $\Delta\theta_p<4^{\circ}$ centered around $\theta_p\approx 70-64^{\circ}$. Other parameters are as in Fig.\ref{Fig1} and Fig \ref{Fig2}.
}
\label{Fig4}
\end{figure}

\subsection{Angular distribution of electrons}
Angular deflection of an electron ($\theta_r$) in the position space is defined as the angle between the laser propagation in $\vekt{z}$ (which is also the direction of $\vekt{B}_{ext} = B_0\vekt{\hat{z}}$) and the transverse plane $x-y$. It is given by
\begin{equation}
\theta_{r} = \tan^{-1}\left(\frac{r_\perp}{z}\right); \,\,\, {\textrm{where}} \,\,\, r_\perp = \sqrt{x^2 + y^2}.
\end{equation}
Figure~\ref{Fig3} shows histograms of electrons vs $\theta_r$ (a1,b1,c1, left column) and 
respective polar plots (a2,b2,c2, right column) with their normalized position $r/R_0$ vs $\theta_r$ corresponding to those energy spectra (A,B,C) in Fig.\ref{Fig2}. Polar co-ordinates ($r,\theta_r$) are color-coded with their energy normalized by $\Up$.   
For lower $B_0=0.02$~a.u. electrons are spread over a wide angular range $\approx 0-175^{\circ}$ (Fig.\ref{Fig3}~a1). Distribution in the ($r,\theta_r$)-plane explains that, the angular spreading contains only low energetic electrons due to weak coupling of laser to the cluster electrons at lower $B_0$ values (Fig.\ref{Fig3}~a2). As the magnetic field increases to $B_0=0.057$~a.u. (Fig.\ref{Fig3}~b1,b2) and $B_0=0.07$~a.u. (Fig.\ref{Fig3}~c1,c2) the electrons align themselves more towards the magnetic field direction $\vekt{z}$ within an angular spreading of $\Delta\theta_r < 5^{\circ}$. This demonstrates that the ambient magnetic field near ECR/RECR probes the ejected electrons to form a collimated beam. For $B_0=0.057$~a.u. and $B_0=0.07$~a.u., energy of most of the electrons in the collimated beam in Figs.\ref{Fig3}~(b2,c2) show maximum absorption satisfy the energy distribution peaks in Figs.\ref{Fig2}~(B,C) respectively. 

The collimation of electrons can also be explained by the angular deflection ($\theta_p$) in the momentum space with transverse momentum ($p_x,p_y$) and longitudinal momentum ($p_z$) as
\vspace{-0.5cm}
\begin{equation}
\theta_{p} = \tan^{-1}\left(\frac{p_\perp}{p_z}\right) ; \,\,\, {\textrm{where}} \,\,\, p_\perp = \sqrt{{p_x}^2 + {p_y}^2}.
\end{equation}
Figure~\ref{Fig4} shows histograms of electrons vs $\theta_p$ (a1,b1,c1, left column) and 
respective polar plots (a2,b2,c2, right column) with their normalized momentum $p/c$ vs $\theta_p$ corresponding to those energy spectra in Fig.\ref{Fig3}. Co-ordinates ($p,\theta_p$) are color-coded with their energy normalized by $\Up$.
For $B_0=0.02$~a.u., there is a wide angular spread centered around $\theta_p\approx 90^{\circ}$, with low energetic electrons in Figs.\ref{Fig4}(a1, a2). However, for higher magnetic fields $B_0=0.057$~(a.u.) in Figs.\ref{Fig4}(b1, b2) and $B_0=0.07$~(a.u.) in Figs.\ref{Fig4}(c1, c2) collimated beams of high energetic electrons are formed in momentum space with angular spreading $\Delta\theta_p < 5^{\circ}$ similar to $\Delta\theta_r < 5^{\circ}$ in the position space. Momentum of beam electrons reach weakly relativistic values 
$p = \sqrt{p_\perp^2 + p_z^2} \approx 0.875 c,1.25 c$ (b2,c2) even with short 5-cycle laser pulse of intensity $\I0\approx 7.13\times 10^{16}\Wcmcm$. 

\section{Effects of cluster size variation}
\label{sec4}
In a realistic scenario, cluster size may vary. The effect of ECR/RECR with an ambient magnetic field on different cluster size is not known. Particularly, it is important to know whether a bigger cluster absorbs more laser energy via ECR/RECR compared to a smaller cluster of $R_0\approx 2.2$~nm in Sec.\ref{sec3}. Therefore we simulate bigger deuterium clusters of $R_0\approx 3.3, 4.4$~nm having number of atoms $N=7208, 17256$ respectively.
However, to accommodate bigger clusters as well as to obtain good accuracy, we now increase the number of computational grids to $128^3$ and the simulation box size to $2048^3$~a.u. in the PIC simulation keeping other simulation parameters/configurations same as the smaller cluster in Sec.\ref{sec3}. 

\begin{figure}[]
\includegraphics[width=0.45\textwidth]{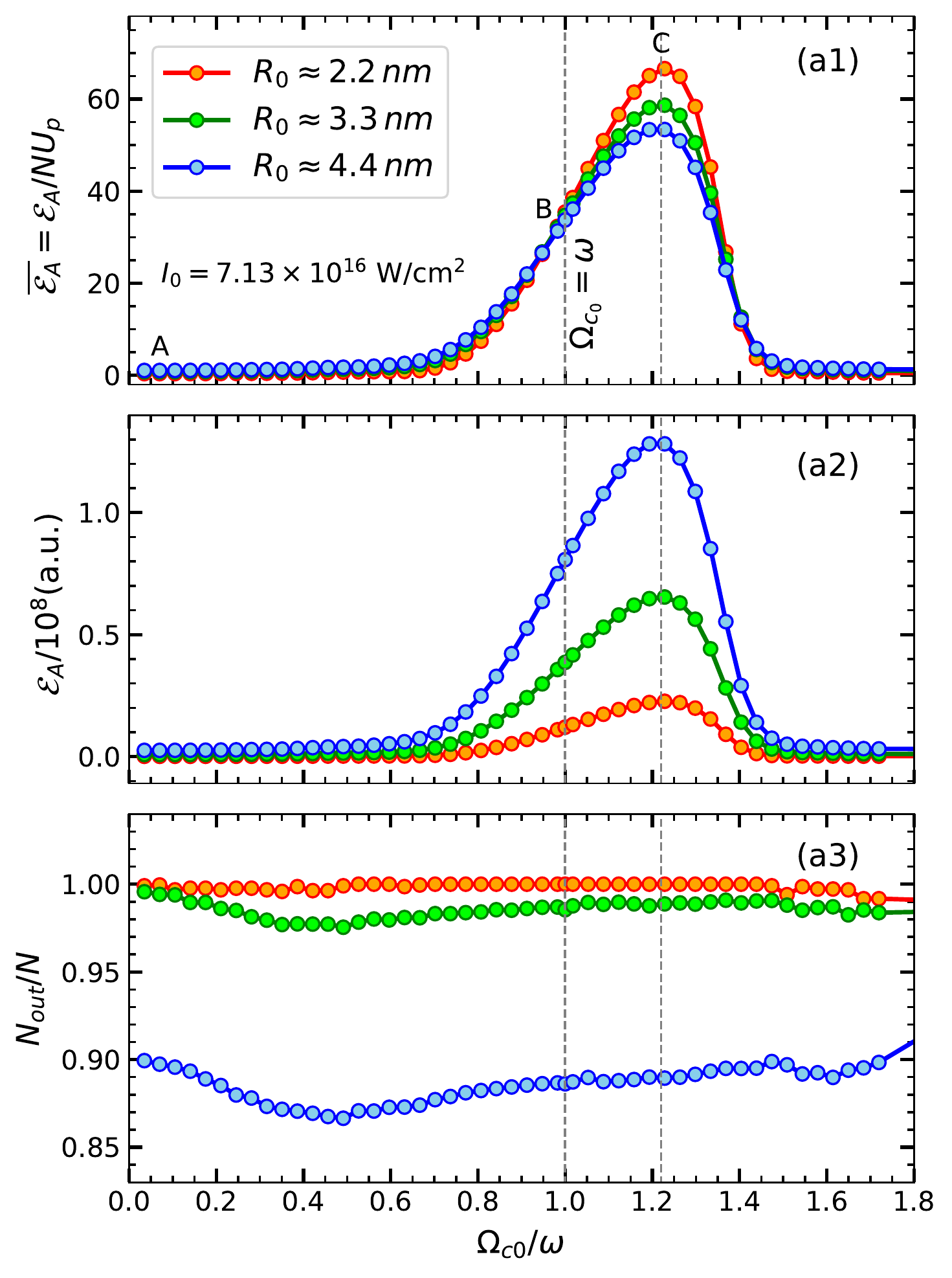}
\vspace{-0.25cm}
\caption{Results for different cluster sizes with $I_0 = 7.13\times10^{16} W/cm^2$: (a1) average total energy ($\overline{\mathcal{E}}_A=\mathcal{E}_A/N \Up$) per cluster-electron in units of $\Up$,  (a2) absorbed total energy (${\mathcal{E}}_A$) in the cluster scaled down by $10^8$, (a3) fractional outer-ionization of electrons ($N_{out}/N$) vs normalized electron-cyclotron frequency $\omegaC0/\omega$ for a range of ambient field $\vert \vekt{\hat{z}} B_{ext}\vert\approx (0 - 2\omega)$ for different cluster sizes $R_0\approx 2.2, 3.3, 4.4$~nm and respective number of atoms $N = 2176, 7208, 17256$. Other parameters are same as in Fig.\ref{Fig1}.
}
\label{Fig5}
\end{figure}

Figures~\ref{Fig5}(a1,a2,a3) show the comparison of absorption per electron $\overline{\mathcal{E}_A} = {\mathcal{E}_A}/N U_p$, total absorption ${\mathcal{E}_A}$ and outer ionized fraction $N_{out}/N$ of electrons vs $\omegaC0/\omega$ of three different cluster sizes for a range of $B_0 = 0-2\omega$ in the end of $n=5$-cycle pulse of intensity $I_0 = 7.13\times10^{16} W/cm^2$. Here $N_{out}$ is the number of freed electrons those have left the cluster boundary. It is evident that increasing cluster size does not significantly affect the absorption peak location (a1,a2) and value of per electron energy $\overline{\mathcal{E}_A}$ upto ECR (a1). The maximum absorption per electron [$\max(\overline{\mathcal{E}_A})$] at the peak gradually drops to $\approx 68\Up,59\Up,54\Up$ as cluster size increases $R_0=2.2, 3.3, 4.4$~nm (a1), which is partly due to relatively less number of outer-ionized electrons ($N_{out}$) for the bigger cluster. 
The outer-ionization of $R_0\approx 4.4$~nm cluster (a3) at this $I_0 = 7.13\times10^{16} W/cm^2$ is $\approx 85-90\%$ which means mainly $85-90\%$ electrons contribute to the total energy absorption. However, for the $R_0\approx 2.2, 3.3$~nm clusters, outer-ionization are around $100\%$ and $98\%$ respectively, indicating nearly all electrons contribute to the energy absorption. Additionally,
restoring force on electrons due to background ions gradually increases as the cluster size increases, which yields relatively higher per electron energy (a1) at the peak $\max(\overline{\mathcal{E}_A})\approx 68\Up$ for a smaller $R_0\approx 2.2$~nm cluster compared to $\max(\overline{\mathcal{E}_A})\approx 48\Up$ for a bigger $R_0\approx 4.4$~nm cluster.
{\em In contrast}, the total energy absorption ${\mathcal{E}_A}$ by electrons (a2) gradually increases with increasing cluster size at a given $B_0$ and for a bigger cluster it is significantly higher at the peak due to more energy carriers ($N_{out}$) after the outer-ionization.
\begin{figure}[]
\includegraphics[width=1.0\linewidth]{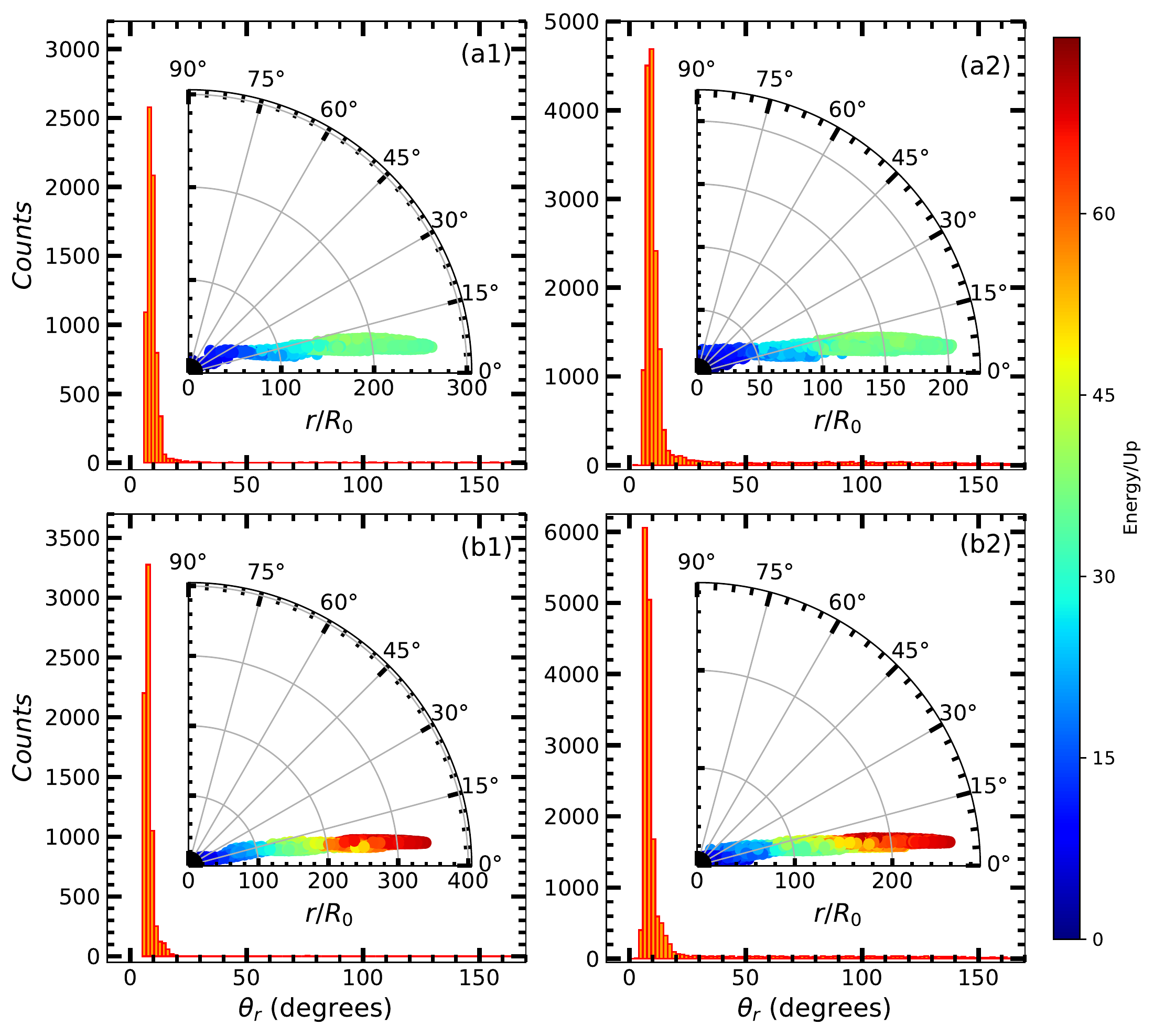}
\vspace{-0.5cm}
\caption{Results for different cluster sizes with $I_0 = 7.13\times10^{16} W/cm^2$:
Histograms/distributions of angular deflection $\theta_r$ of PIC electrons and 
respective polar plots (insets) with their $r/R_0$ vs $\theta_r$ corresponding to data points at (B,C) in Fig.\ref{Fig5}a1 with $B_0=0.057, 0.07$~a.u. for $R_0\approx 3.3$~nm cluster with $N = 7208$ atoms (a1,b1, left column) and $R_0\approx4.4$~nm cluster with $N = 17256$ atoms (a2,b2, right column) respectively. Polar co-ordinates ($r,\theta_r$) of electrons (insets) are color-coded with their energy normalized by $\Up$. Ejected electrons propagate long distance $r = \sqrt{r_\perp^2 + z^2} \approx 350 R_0,280 R_0$ (b1,b2) as collimated beams with angular spreads $\Delta\theta_r<3^{\circ}$ centered around $\theta_r\approx 7-8^{\circ}$. Other parameters are same as in Fig.\ref{Fig1}. 
}
\label{Fig6}
\end{figure}
For many applications, higher flux of energetic electrons as a collimated beam may be required and bigger clusters may supply them. Therefore, in Fig.\ref{Fig6} we plot histograms for angular distribution of electrons and corresponding polar plots (insets) for $R_0=3.3, 4.4$~nm clusters (as in Fig.\ref{Fig3} with 2.2~nm cluster) corresponding to $B_0\approx 0.057$~a.u. (at B in Fig.\ref{Fig5}a1) and $B_0\approx 0.07$~a.u. (at C in Fig.\ref{Fig5}a1) respectively. Compared to $2.2$~nm cluster (Fig.\ref{Fig3}), the angular spread $\Delta\theta_r$ is little wider for larger clusters (Fig.\ref{Fig6}), but there are now more number of energetic electrons within $\theta_r\approx 8^{\circ}-12^{\circ}$. In the case of $3.3$~nm cluster for $B_0=0.057, 0.07$~a.u. (a1,b1, left) the range of $\theta_r$ is almost same, however the electron beam with $B_0=0.07$~a.u. contains higher number ($\approx 3300$) of energetic electrons around $\theta_r\approx 8^{\circ}$. Similarly for $4.4$~nm cluster, due to its bigger size the energetic electron population in the collimated beam increases to $\approx 5000, 6000$ for $B_0=0.057, 0.07$~a.u. (a2,b2, right) around $\theta_r\approx 8^{\circ}$. We may conclude that the collimated electron beams become more intense with greater number of energetic electrons as cluster size increases which may not be possible without ambient $B_{0}$.
\subsection{Effects at high intensity}
\label{sec5}
\begin{figure}[]
\includegraphics[width=0.45\textwidth]{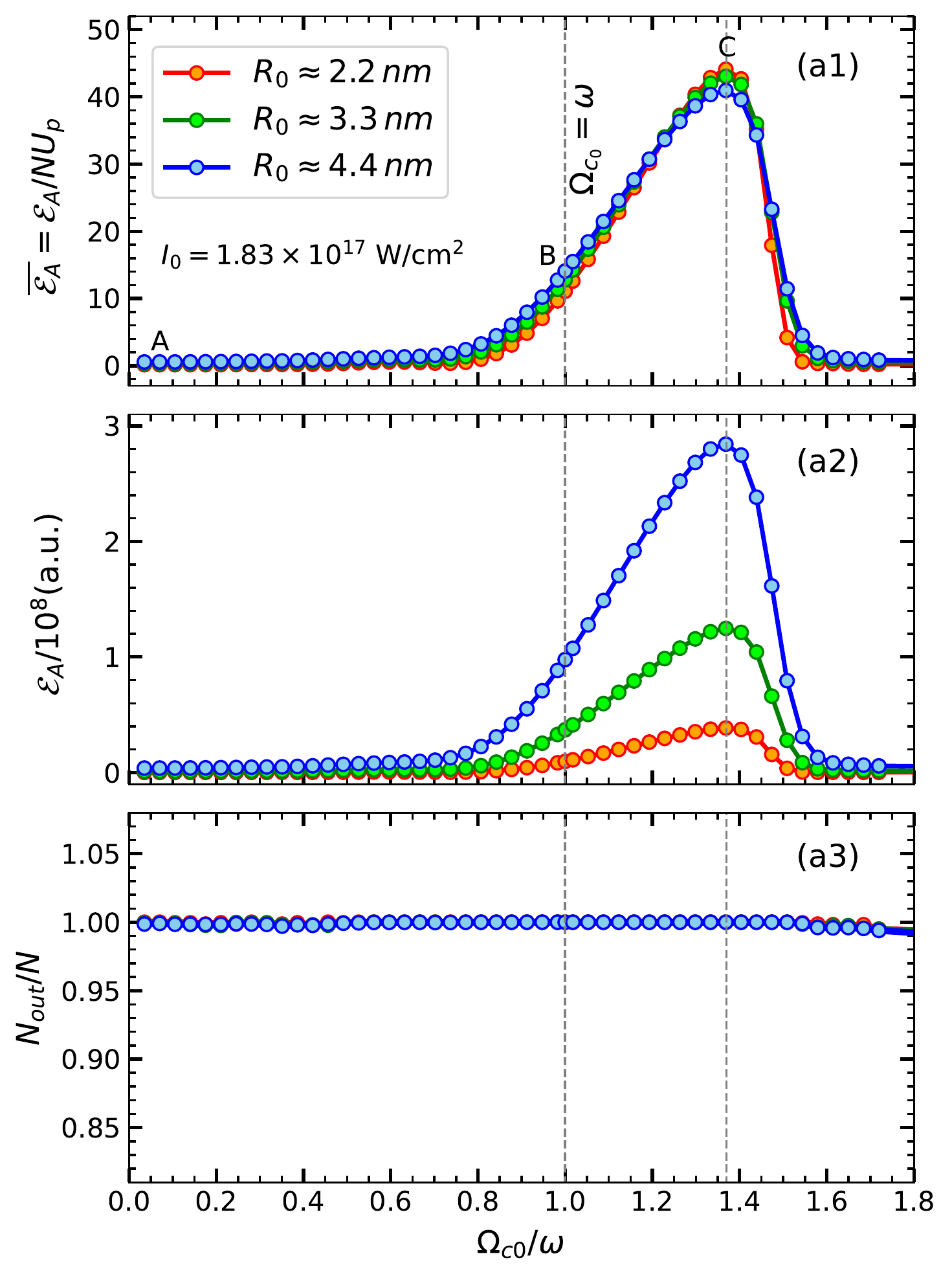}
\vspace{-0.25cm}
\caption{ Results for different cluster size with $I_0 = 1.83\times10^{17} W/cm^2$: (a1) average total energy ($\overline{\mathcal{E}}_A=\mathcal{E}_A/N \Up$) per cluster-electron in units of $\Up$,  (a2) absorbed total energy (${\mathcal{E}}_A$) in the cluster scaled down by $10^8$, (a3) fractional outer-ionization of electrons ($N_{out}/N$) vs normalized electron-cyclotron frequency $\omegaC0/\omega$ for a range of ambient field $\vert \vekt{\hat{z}} B_{ext}\vert\approx (0 - 2\omega)$ for different cluster sizes $R_0\approx 2.2, 3.3, 4.4$~nm and respective number of atoms $N = 2176, 7208, 17256$. Note that $I_0 = 1.83\times10^{17} W/cm^2$ corresponds to greater $U_p = 402.26$~a.u. compared to Fig.\ref{Fig5}. Other parameters are same as in Fig.\ref{Fig1}.
} 
\label{Fig7}
\end{figure}

Results in previous sections are obtained with $I_0 = 7.13\times10^{16} W/cm^2$. In case of 4.4~nm cluster (Fig.\ref{Fig5}a3) nearly 10-15\% electrons are still within the cluster at this intensity. 
We now perform PIC simulations at a higher $I_0 = 1.83\times10^{17} W/cm^2$ (still in the non-relativistic regime) for all three cluster sizes $R_0=2.2, 3.3, 4.4$~nm, keeping other parameters same as in Fig.\ref{Fig5}. 
Results of absorption per electron $\overline{\mathcal{E}_A} = {\mathcal{E}_A}/N U_p$ in $\Up$, total absorption ${\mathcal{E}_A}$ and outer ionized fraction $N_{out}/N$ of electrons vs $\omegaC0/\omega$ are shown in Fig.\ref{Fig7}. 
Compared to Fig.\ref{Fig5}, per electron energy
$\overline{\mathcal{E}_A}$ for ejected electrons from different clusters are nearly same (Fig.\ref{Fig7}a1) and maximum absorption per electron now ranges $\overline{\mathcal{E}_A}\approx 44-42\Up$ as cluster size increases $R_0=2.2-4.4$~nm, but outer-ionization reaches 100\% for all clusters (Fig.\ref{Fig7}a3) at this higher intensity.  
Total absorption ${\mathcal{E}_A}$ in each cluster (Fig.\ref{Fig7}a2) increases more than two times than in Fig.\ref{Fig5}a2. If we compare the ratio of maximum absorption (Fig.\ref{Fig7}a2) for different clusters, we find $\max({\mathcal{E}_A})\approx 0.4:1.3:2.8\approx 1: 3.325: 6.9$ which scales with the number electrons in the cluster as $N \approx 2176:7208:17256 \approx 1:3.3:7.9$. Thus $N$ vs $\max({\mathcal{E}_A})$ is almost linear at a very high $I_0$ when outer-ionization is $100\%$.

\begin{figure*}[]
\includegraphics[width=0.75\textwidth]{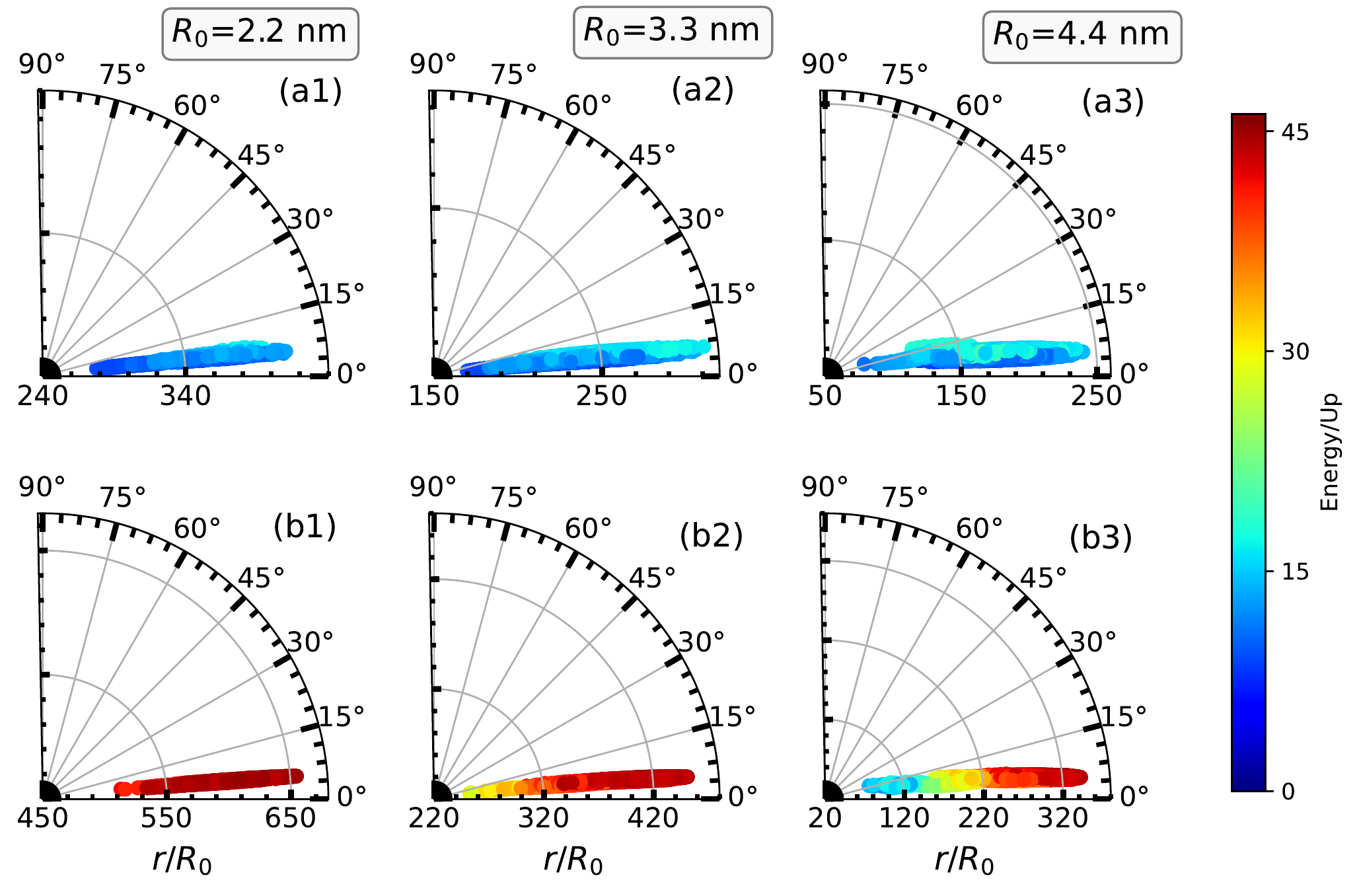}
\vspace{-0.5cm}
\caption{Results for different cluster size (column wise) with higher $I_0 = 1.83\times10^{17}\, \Wcmcm$:
Angular distribution of PIC electrons  
in the polar-plane with $r/R_0$ vs $\theta_r$ corresponding to those data points at (B,C) in Fig.\ref{Fig7} for $B_0=0.057, 0.078$~a.u. (top,bottom) respectively. Panels (a1,b1, left column) for 2.2~nm cluster (2176 atoms), (a2,b2, middle column) for 3.3~nm cluster (7208 atoms) and (a3,b3, right column) for 4.4~nm cluster (17256 atoms). Polar co-ordinates ($r,\theta_r$) of electrons are color-coded with their energy normalized by $\Up$. Ejected electrons propagate long distance beyond $r = \sqrt{r_\perp^2 + z^2} \approx 650 R_0, 430 R_0, 330 R_0$ (b1,b2,b3) as collimated beams with angular spreads $\Delta\theta_r<3^{\circ}$ centered around $\theta_r\approx 6-7^{\circ}$. Other parameters are same as in Fig.\ref{Fig1} and Fig.\ref{Fig2}. 
}
\label{Fig8}
\end{figure*}
In Fig.\ref{Fig8} we compare the distribution of ejected electrons in ($r,\theta_r$) plane (as in Figs.\ref{Fig3},\ref{Fig6}) for three different cluster sizes (column wise) corresponding to points (B,C) in Fig.\ref{Fig7} with $B_0=0.057, 0.078$~a.u. (top,bottom). We find respective electron beams are even more collimated within an angular range of $\Delta \theta_r \approx 3^\circ - 4^\circ$. Also, with increasing cluster size, electron beams are more intense with greater number of energetic electrons. At a very high $I_0$ and ambient $B_0$ near ECR/RECR, electron distribution for a bigger cluster becomes very similar to that of a small cluster in the regime of 100\% outer-ionization. 

\vspace{-0.5cm}
\section{Conclusion}
\label{sec6}
\vspace{-0.25cm}
We study interaction of intense 800~nm, 5-fs (fwhm) broadband laser pulses of different intensities $I_0 = 7.13\times 10^{16} - 1.83\times 10^{17}\, \Wcmcm$ with deuterium clusters of various sizes (radius $R_0\approx 2.2-4.4$~nm) in presence of ambient magnetic fields of strengths $B_0 =0-2\omega$ along the laser propagation direction $\vekt{z}$ using 3D hybrid-PIC simulations.  
Here laser absorption occurs in two stages via AHR ($1^{st}$ stage) and electron cyclotron resonance (ECR) or relativistic ECR (RECR) processes ($2^{nd}$ stage). Auxiliary $B_0$ enhances coupling of laser to cluster-electrons via improved frequency-matching for ECR/RECR as well as phase-matching\cite{Swain2022} for prolonged duration of the pulse (which are also checked for bigger clusters, but not repeated here for conciseness) and the average absorbed energy per electron $\overline{\mathcal{E}}_A$ jumps to $\overline{\mathcal{E}}_A \approx 36-70\,\Up$ which {\em is significant}. Otherwise $ \overline{\mathcal{E}}_A$ is mostly limited around $\overline{\mathcal{E}}_A\approx  0.5 - 3\Up$ without $B_0$. Increasing the cluster size ($R_0\approx 2.2 \rightarrow 4.4$~nm) per electron energy $\overline{\mathcal{E}}_A$ remains almost same ($\overline{\mathcal{E}}_A \approx 36-70\,\Up$) near ECR/RECR, but net absorption increases almost linearly with number of electrons $(N)$ in the regime of $100\%$ outer-ionization at high intensities.

We further analyze the energy distribution of ejected electrons as well as their angular distribution in the position space and in the momentum space. We find that laser coupled electrons form a nearly mono-energetic, weakly relativistic collimated beam that traverse a few hundreds of $R_0$ (or on the order of $\lambda$) in presence of an ambient magnetic field near ECR/RECR which may not be possible only with the laser field. Also, as the cluster size increases, intensity of electron beam increases with greater number of energetic electrons at a restricted angle of $\theta_r \approx 7^\circ - 10^\circ$ w.r.t. $\vekt{z} $ direction for $I_0 = 7.13\times10^{16} - 1.83\times10^{17} \Wcmcm$. 

This work may find importance for the fast ignition technique of inertial confinement fusion where intense collimated relativistic electron beam (REB) is required to be transported deep inside the matter with less divergence, laser-driven electron accelerators, ultra-short x-ray sources for radiation therapy and other medical applications.

\section*{Acknowledgements}
Numerical simulation works have been performed in Antya Linux cluster of HPC facility at IPR. Authors acknowledge Dr. Devendra Sharma for careful reading of the manuscript.

\bibliography{MSClusterKSMK}
\end{document}